\documentclass[cameraready]{Interspeech}

\title{Confidence Score Guided Incremental and Speaker Adaptive Pseudo-Labeling for Semi-Supervised Elderly Speech Recognition}

\author[affiliation={1}]{Chengxi}{Deng}
\author[affiliation={2*}]{Xurong}{Xie}
\author[affiliation={1}]{Shujie}{Hu}
\author[affiliation={1}]{Jiajun}{Deng}
\author[affiliation={3}]{Mengzhe}{Geng}
\author[affiliation={1}]{Youjun}{Chen}
\author[affiliation={1}]{Huimeng}{Wang}
\author[affiliation={1}]{Haoning}{Xu}
\author[affiliation={1}]{Guinan}{Li}
\author[affiliation={1*}]{Xunying}{Liu}

\address{
    $^1$ The Chinese University of Hong Kong, Hong Kong SAR, China \\
    $^2$ Institute of Software, Chinese Academy of Sciences, China \\
    $^3$ National Research Council Canada, Canada 
}

\email{\{cxdeng, xyliu\}@se.cuhk.edu.hk, xurong@iscas.ac.cn}


\usepackage{comment}
\usepackage{amsmath}
\usepackage{algorithm}
\usepackage{algorithmic}

\usepackage{multirow}
\usepackage{pifont}
\newcommand{\cmark}{\ding{51}}
\newcommand{\xmark}{\ding{55}}
\usepackage{graphicx}
\usepackage[table]{xcolor}
\usepackage{bm}
\usepackage{amsmath,amsfonts,amssymb,mathtools}
\usepackage{cite}
\usepackage[utf8]{inputenc}
\usepackage[titletoc,title]{appendix}
\usepackage{makecell}
\usepackage{tipx}
\usepackage{color}
\usepackage{multirow}
\usepackage{hhline}

\begin{document}

\maketitle
\renewcommand{\thefootnote}{*}%
\footnotetext{Corresponding author.}%
\renewcommand{\thefootnote}{\arabic{footnote}}%

\begin{abstract}

This paper proposes a novel confidence score guided incremental and speaker adaptive pseudo-labeling approach for semi-supervised elderly speech recognition. It facilitates higher-quality pseudo-label selection and progressive refinement, while also mitigating speaker heterogeneity. A confidence estimation module is designed to rank the reliability of untranscribed data, enabling a curriculum learning trajectory that progressively folds in unlabeled data subsets from high to low confidence. Speaker-specific characteristics are captured through speaker adaptive training with learnable prompts. Experiments on the English DementiaBank Pitt and Cantonese JCCOCC MoCA elderly speech datasets suggest that the proposed method outperforms the semi-supervised baseline using no confidence scores guided incremental or speaker adaptive pseudo-labeling by statistically significant word error rate (WER) or character error rate (CER) reductions of 1.45\% and 2.27\% absolute (6.21\% and 6.98\% relative).

\end{abstract}

\section{Introduction}
In the context of global population aging, ensuring effective communication for the elderly becomes increasingly vital for maintaining their social participation and quality of life. Elderly speech exhibits significant heterogeneity, including varying degrees of imprecise articulation stemming from weakened neuromotor control and linguistic degradation associated with cognitive decline\cite{becker1994natural_dbank}. Since current foundation models primarily target normal speakers\cite{whisper,baevski2020wav2vec,hsu2021hubert,chen2022wavlm,conneau21_interspeech}, their application to elderly speech remains a challenging task\cite{hu2022exploring,ssl_shujie_taslp,deng25_interspeech}. Therefore, it is essential to investigate how to effectively adapt these foundation models to better accommodate elderly speech. Concurrently, the rapid expansion of remote healthcare and tele-consultation services has led to the accumulation of substantial unlabeled elderly speech data, creating an urgent need for methods that enable foundation models to learn effectively from such unlabeled resources.\par
Elderly speech presents multifaceted challenges to existing deep learning-based ASR technologies, including:  {\textbf{1) Labeled data scarcity} \cite{geng2022speaker,geng2024homogeneous,Personalized_data_aug}:} The accurate transcription of pathological speech necessitates clinical expertise, leading to significant labeling costs and a shortage of annotated data. {\textbf{2) Untrustworthy auto-labeling}~\cite{ssl_shujie_taslp,deng25_interspeech}:} Advanced speech foundation models fine-tuned on labeled in-domain elderly speech data still exhibit high word error rates. Directly employing such models to decode unlabeled speech often yields unreliable pseudo-labels that hinder the effective utilization of unlabeled data. {\textbf{3) Speaker heterogeneity}~\cite{geng2022speaker,geng2024homogeneous}:} Among elderly speakers, typical sources of variability, such as accent and gender, are further compounded by varying degrees of phonetic deterioration and linguistic degradation. Such heterogeneity complicates the training of speaker-independent (SI) ASR systems, further degrading the quality of pseudo-labels during decoding and limiting the effectiveness of semi-supervised learning approaches.\par

Semi-supervised learning has proven effective across different generations of ASR systems and has been successfully extended to advanced foundation models for normal speech~\cite{likhomanenko21b_interspeech,park25_interspeech,zheng2026pseudo,damianos25_interspeech,park20d_interspeech,hwang22c_interspeech}. Most of these approaches adopt a pseudo-labeling strategy, where a model trained on labeled data is used to decode unlabeled utterances, with the generated pseudo-labels then serving as supervision for further training. Despite their effectiveness, such methods carry inherent risks, as unreliable pseudo-labels can degrade model performance~\cite{ZhuGCPZY23}. To mitigate sensitivity to unreliable labels, researchers have proposed confidence-based filtering that discards low-confidence pseudo-labels~\cite{rangappa25_interspeech,Large_Scale_self_semi}, as well as incremental training strategies that partition pseudo-labels into subsets and iteratively refine them through model updates~\cite{carofilis25_interspeech,xu20b_interspeech}.

However, these prior studies suffer from the following limitations when directly performed on the elderly speech:  \textbf{a)} Confidence-based filtering assumes that low-confidence outputs correspond to unreliable pseudo-labels and discards them entirely, neglecting to improve pseudo-label quality~\cite{rangappa25_interspeech,Large_Scale_self_semi}. \textbf{b)} Prior incremental methods partition pseudo-labels into subsets without considering pseudo-label reliability~\cite{carofilis25_interspeech,xu20b_interspeech}. When low-quality pseudo-labels appear in early training stages, the corrupted model updates degrade the quality of regenerated pseudo-labels for subsequent subsets, leading to error accumulation throughout the iterative process. \textbf{c)} Existing approaches ~\cite{rangappa25_interspeech,Large_Scale_self_semi,carofilis25_interspeech,xu20b_interspeech} directly employ speaker-independent (SI) models for decoding, without leveraging speaker modeling which has been shown effective in mitigating the heterogeneity among elderly speakers~\cite{hu2024structured,deng25_interspeech,geng2022speaker,geng2024homogeneous}, resulting in lower-quality pseudo-labels.

\begin{figure*}[h]
    \centering    
    \includegraphics[width=0.8\textwidth]{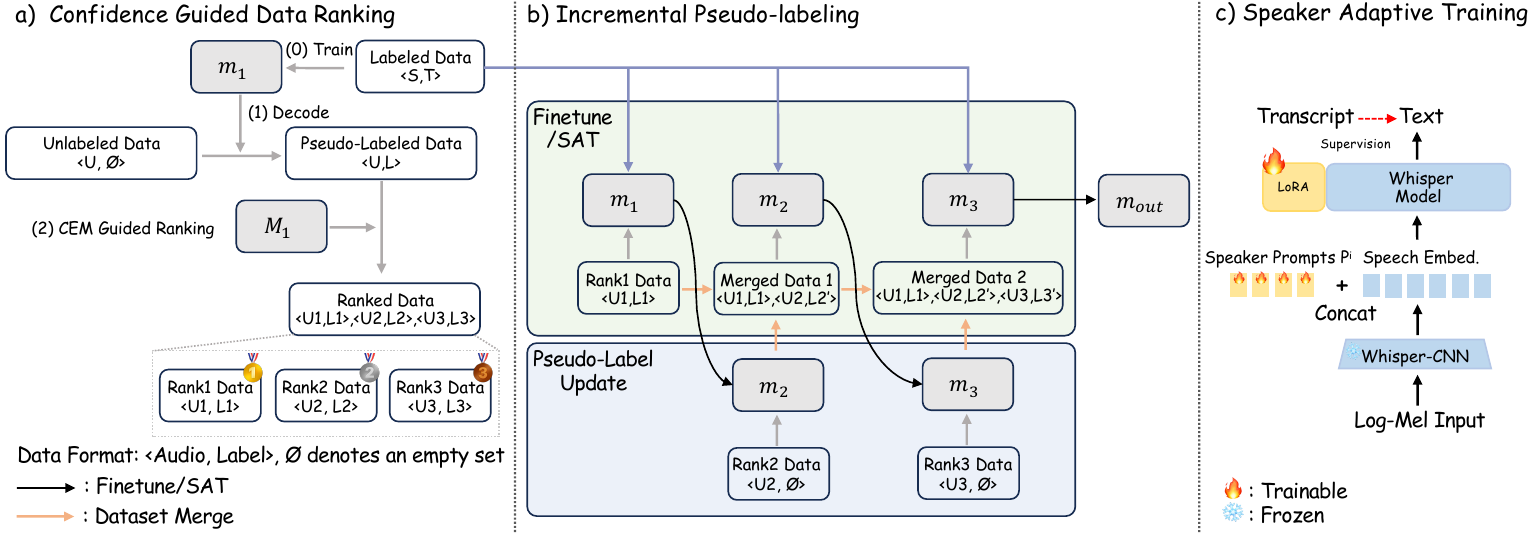}
    \caption{Examples of confidence score guided pseudo-labeling with 3 iterations (a) \& b)) and speaker adaptive training (c)).}
\label{pipeline_method}
\vspace{-0.7cm}
\end{figure*}

To address the sensitivity to label errors in semi-supervised model fine-tuning, a lightweight confidence score estimation module (CEM) is purpose-designed for Whisper to produce a ranking order of trustworthiness for its error-prone decoding outputs of untranscribed elderly speech training data. This allows both \textbf{1)} more reliable portions of the unlabeled training data and their Whisper ASR transcripts to be selected for semi-supervised model fine-tuning together with small quantities of labeled data; and \textbf{2)} progressive transcription generation and semi-supervised model fine-tuning by gradually folding in subsets of unlabeled training data that are ranked by confidence scores in a descending order, akin to a curriculum learning trajectory\cite{bengio2009curriculum} steadily ramping up in task difficulty. Speaker prompt-adapted Whisper models are further employed to produce more accurate speech transcripts, providing a robust foundation for unsupervised test-time adaptation.

The main contributions of this work are as follows:\\
\textbf{1)} To the best of our knowledge, this paper presents the first study on confidence-guided incremental and speaker adaptive pseudo-labeling for semi-supervised elderly speech recognition. This can be summarized in three key points: \textbf{a)} Compared with confidence-based filtering methods that discard low-confidence samples~\cite{rangappa25_interspeech,Large_Scale_self_semi}, our method progressively regenerates higher-quality pseudo-labels during model updates, incrementally improving pseudo-label quality across stages. \textbf{b)} Compared with prior incremental methods that partition pseudo-labels without considering their reliability~\cite{carofilis25_interspeech,xu20b_interspeech}, we propose a confidence-guided curriculum strategy that processes samples from high to low confidence, mitigating error accumulation throughout the iterative training process. \textbf{c)} Unlike previous semi-supervised approaches that employ speaker-independent models~\cite{rangappa25_interspeech,Large_Scale_self_semi,ZhuGCPZY23,carofilis25_interspeech,xu20b_interspeech}, our method leverages speaker modeling to capture speaker-specific characteristics, facilitating higher-quality pseudo-label generation. The resulting speaker-adapted model also provides a robust foundation for unsupervised test-time adaptation to unseen speakers.\\
\textbf{2)} Experimental results on the DementiaBank Pitt \cite{becker1994natural_dbank} and JCCOCC MoCA \cite{jccocc_datasets} elderly speech datasets suggest that the proposed method outperforms the semi-supervised baseline by statistically significant WER or CER reductions of \textbf{1.45\%} and \textbf{2.27\%} absolute (\textbf{6.21\%} and \textbf{6.98\%} relative).
\vspace{-0.3cm}
\section{Large-Scale Foundation Model Whisper}
Whisper is a transformer-based multilingual speech foundation model that takes log-Mel spectrogram $\bm{X} \in \mathbb{R}^{D \times F}$ as input, where $D$ and $F$ indicate the dimension and length, respectively. The input is downsampled by a factor of 1/2 through the convolutional block, and the encoder maps it to the encoder feature. The decoder then generates the next token conditioned on the encoder feature, previous tokens and special tokens.

\begin{table*}[htbp]
\caption{Performance contrast of the proposed confidence-based incremental speaker modeling and different comparable semi-supervised methods on DementiaBank Pitt and JCCOCC MoCA. ``Sup." and ``Semi-Sup." denote whether the training process uses only labeled data or a mixture of labeled data with corresponding pseudo-label data, respectively. ``Inv." and ``Par." refer to clinical investigators and elderly participants. $^\ast$ denotes statistically significant (MAPSSWE \cite{gillick1989some}, $\alpha$ = 0.05) improvements obtained against the semi-supervised baseline ASR systems (Sys.3)}
    \centering
    \vspace{-0.35cm}
    \resizebox{\linewidth}{!}
    {
    \begin{tabular}{c|c|c|c|c|c|c|cc|cc|c|c|c|c}
    \hline\hline 
    \multirow{3}{*}{Sys.} & 
    \multicolumn{4}{c|}{\shortstack{Pseudo Labeling}} & 
    \multirow{3}{*}{\shortstack{TTA}} & 
    \multirow{3}{*}{\shortstack{Finetune\\Strategy}} & 
    \multicolumn{5}{c|}{DementiaBank Pitt WER(\%)} &
    \multicolumn{3}{c}{JCCOCC MoCA CER(\%)} \\
    \cline{2-5}
    \cline{8-15}
    &\multirow{2}{*}{\shortstack{Percentile of\\Data Used}} &
    \multirow{2}{*}{\shortstack{Data\\Selection}} & 
    \multirow{2}{*}{\shortstack{Incremental\\Iteration}} & 
    \multirow{2}{*}{SAT}&&&
    \multicolumn{2}{c|}{Dev.}&\multicolumn{2}{c|}{Eval.} & \multirow{2}{*}{All}&
    \multirow{2}{*}{Dev.} & \multirow{2}{*}{Eval.}
    & \multirow{2}{*}{All}\\
    \cline{8-11}
    & & & & & & &Par.&Inv.&Par.&Inv.&&&&\\
    \hline\hline    

1&-&-&-&-&-&100\% Sup.&28.79 &12.76&20.68&12.65&20.43&28.68&25.79&27.23\\
\cline{2-15}

{\cellcolor{yellow!25}2}&-&\multirow{2}{*}{\xmark}&\multirow{3}{*}{\xmark}&\multirow{5}{*}{\xmark}&\multirow{5}{*}{\xmark}&10\% Sup.&33.84&16.38&23.70&13.43&24.43&35.39&31.84&33.60 \\

{\cellcolor{green!10}3}&100\%&&&&& Semi-Sup.&32.03&15.57&23.62&13.32&23.36&34.17&30.86&32.52 \\
\cline{2-3}

4&\multirow{3}{*}{80\%}&Conf. Score&&&&Semi-Sup.&31.64&15.07&22.40&13.54&22.81$^\ast$&33.84&30.50&32.16 \\
\cline{4-4}

5&&Random&\multirow{2}{*}{4}&&&Semi-Sup. &31.71&15.34&22.49&14.54&22.99&33.69&30.46&32.06 \\

{\cellcolor{cyan!15}6}&&Conf. Score&&&&Semi-Sup.&30.85$^\ast$&14.91&22.23$^\ast$&14.87&22.45$^\ast$&33.15$^\ast$&29.84$^\ast$&31.48$^\ast$\\
\hline\hline

7&-&-&\multirow{2}{*}{\xmark}&\multirow{3}{*}{\cmark}&\multirow{3}{*}{\cmark}&10\% Sup.&32.44&15.84&22.86&13.32&23.51&34.89&31.36&33.11\\ 
\cline{2-3}
8&\multirow{2}{*}{80\%}&\multirow{2}{*}{Conf. Score}&&&& Semi-Sup.&31.50&14.99&22.13$^\ast$&12.87&22.66$^\ast$&33.63&30.17&31.89\\ 
\cline{4-4}
{\cellcolor{orange!20} 9}&&&4&&&Semi-Sup.&30.26$^\ast$&14.58$^\ast$&21.50$^\ast$&13.32&21.91$^\ast$&31.53$^\ast$&28.98$^\ast$&30.25$^\ast$\\ 
\hline\hline
\end{tabular}
}
\vspace{-0.7cm}
\label{tab:main_results}
\end{table*}

\vspace{-0.3cm}
\section{Confidence Score Guided Data Selection}
\par
\noindent
\textbf{Confidence Score Estimation:} 
Directly employing models trained on limited labeled data to decode unlabeled elderly speech often yields unreliable pseudo labels. This motivates the need for a confidence estimation mechanism that quantifies the reliability of pseudo labels. A straightforward method is to use the decoder output probabilities of the end-to-end model. However, these have been found to suffer from an overconfidence issue~\cite{li2021confidence,liu2021utterance,hu2024self}. To address this issue, a lightweight token-level confidence estimation module (CEM) is adopted in this paper to produce more reliable confidence scores.\par
The CEM is a lightweight binary classifier based on a 3-layer residual FFN. It concatenates the decoder output of each token with the top-10 score logits, and outputs a confidence score for the token. During training, the tokens from the hypothesized transcription and the ground truth reference are aligned via edit distance computation, where correctly matched tokens are assigned a target label of one, while substituted or inserted tokens are assigned a label of zero. When estimating the confidence of pseudo labels for unlabeled data, the utterance-level score is obtained by averaging the token-level confidence scores across all tokens within an utterance except special tokens.

\par
\noindent
\textbf{Data Selection:} 
We develop a confidence-based filtering baseline using the CEM, which is initially trained on the labeled data. This approach ranks the pseudo-labels according to their utterance-level confidence in descending order and retains only a specified top percentile. The selected high-confidence pseudo-labeled data is then combined with the labeled data for training.

\vspace{-0.3cm}
\section{Confidence Score Guided Pseudo Labeling}

We propose a novel confidence-guided data ranking strategy combined with an incremental pseudo-labeling mechanism. Conventional confidence-based filtering directly operates on all pseudo-labels, which risks completely discarding utterances from specific elderly speakers, particularly those with severe articulation or language organization issues, thereby further impairing our subsequent speaker modeling. Therefore, we consider ranking and partitioning the data within each speaker to ensure all speakers are preserved. Formally, we define the dataset format as $\langle Audio, Label \rangle$. Labeled and unlabeled sets are denoted as $\langle S, T \rangle$ and $\langle U, \varnothing \rangle$, respectively, where $\varnothing$ denotes an empty set.
\par
\noindent
\textbf{Confidence Guided Data Ranking:} As shown in Fig. 1(a), we finetune the original Whisper model and confidence estimation model on labeled dataset $\langle S, T \rangle$, obtaining $m_1$ and $M_1$, respectively. The model $m_1$ then decodes $\langle U, \varnothing \rangle$ to produce initial pseudo-labeled data $\langle U, L \rangle$. The $M_1$ estimates utterance-level confidence scores. For each speaker, unlabeled utterances are ranked by these scores in descending order and equally partitioned into K groups within that speaker's data. Utterances with the same within-speaker rank are then grouped across speakers to form K subsets $\langle U_1, L_1 \rangle, \langle U_2, L_2 \rangle, \ldots, \langle U_K, L_K \rangle$, (i.e., Rank $1$ to Rank $K$ data), from highest to lowest confidence.
\par
\noindent
\textbf{Incremental Pseudo-Labeling:} As shown in Fig. 1(b), the training proceeds iteratively from the highest-confidence subset to the lowest, following the curriculum learning trajectory. First, $m_1$ is finetuned on the labeled data $\langle S, T \rangle$ combined with $\langle U_1, L_1 \rangle$, yielding $m_2$. For subsequent iterations ($i = 2, 3, \ldots, K$), model $m_i$ regenerates pseudo-labels for subset $i$, transforming $\langle U_i, \varnothing \rangle$ into $\langle U_i, L_i' \rangle$. The relabeled subset $\langle U_i, L_i' \rangle$ is then merged with the labeled data $\langle S, T \rangle$ and all previously accumulated pseudo-labeled subsets $\langle U_1, L_1 \rangle,\langle U_2, L_{2}' \rangle, \ldots, \langle U_{i-1}, L_{i-1}' \rangle$ to train and obtain $m_{i+1}$. The incremental pseudo-labeling strategy improves model performance at early training stages using more reliable pseudo-labels, which in turn generates more accurate pseudo-labels for subsequent subsets. \par

We denote the above pipeline as the \textbf{incremental SI} system. To further enhance the effectiveness of this SI incremental framework, we consider the substantial speaker variability in elderly speech by incorporating speaker modeling into the iterative updates. Specifically, we adopt the advanced speaker prompt adaptation~\cite{deng25_interspeech}, which is detailed in the next section.

\vspace{-0.3cm}
\par
\noindent
\section{Semi-supervised Speaker Adaptive Modeling}
\par
\noindent
\textbf{Speaker Adaptive Training:} For each training speaker, we initialize a set of trainable speaker prompts and concatenate them with the speech features. Considering there are $I$ training speakers, the process can be conducted as follows: 
\vspace{-0.2cm}
\begin{equation}
\begin{aligned}
\bm{H}_{conv}^{i} &= \text{Concat}[\bm{R}^i, {\text{Conv}(\bm{X}^{i})}]
\end{aligned}
\end{equation}
where $i \in \{1,2,...,I\}$ is the index of training speakers, $\bm{R}^{i}$ is the speaker prompts of training speaker $i$, $\bm{X}^i$ denotes the log-Mel spectrogram input of speaker $i$, $\bm{H}_{conv}^i\in \mathbb{R}^{D\times (Q+F/2)}$ is the resulting hidden states obtained by concatenating with the speaker prompts, $Q$ indicates the speaker prompts length. We then perform speaker adaptive training (SAT)~\cite{anastasakos1996compact_sat}, the process can be conducted as follows:  
\vspace{-0.2cm}
\begin{equation}
\{\bm{\hat{R}}^i, \hat{\bm{\Theta}}\} = \underset{\{\bm{R}^{i}, \bm{\Theta}\}}{\arg\min} \{\mathcal{L}_{CE} (\bm{Y}^{i}|\bm{X}^i;\bm{R}^{i}, \bm{\Theta})\}
\end{equation}
where $\bm{\Theta}$ represents the LoRA parameters shared among all training speakers, $\bm{Y}^{i}$ represents the transcript, and $\mathcal{L}_{CE}$ is the cross entropy loss. 
Building upon the incremental SI system, we replace the finetuning step in each iteration with SAT, and utilize the obtained speaker prompts to decode the subsequent subset of data. We denote this process as the \textbf{incremental SAT} system. Compared with SI models, SAT better captures speaker-specific characteristics of training speakers via speaker prompts, which produce higher-quality pseudo-labels when decoding progressively lower confidence data. The resulting speaker-invariant canonical model further provides a robust foundation for unsupervised test-time adaptation.

\par
\noindent
\textbf{Adaptation Data Accumulation:} To estimate speaker-dependent (SD) parameters for test speakers, the corresponding SI system must first decode the test speaker's data and generate pseudo-labels as supervision.
\par
\noindent
\textbf{Test-Time Adaptation:} During test-time adaptation (TTA), a set of learnable speaker prompts is introduced for each test speaker and optimized using pseudo-labels produced by the corresponding SI system. As an optional extension, this procedure can be applied during incremental SAT pipeline, where speaker prompts are optimized using previously decoded pseudo-labels to produce refined pseudo-labels for subsequent SAT, referred to as \textbf{speaker adaptive relabeling (SA relab.)}.

\vspace{-0.3cm}
\section{Experiments}
\par
\noindent
\textbf{Task description:}
The DementiaBank Pitt corpus contains 33 hours of audio from 292 AD assessment interviews. The training set consists of 688 speakers, expanding to 58.9 hours after silence stripping and data augmentation \cite{cuhk_elderly_zi_ye}. The development and evaluation sets include 119 and 95 speakers, each with 2.5 and 0.6 hours of audio. The JCCOCC MoCA corpus contains 32.4 hours of audio from 256 cognitive assessment interviews. The training set consists of 369 speakers, expanding to 156.9 hours after silence stripping and data augmentation \cite{cuhk_elderly_zi_ye}. The development and evaluation sets include 49 speakers, each with 3.5 and 3.4 hours of audio. \textbf{No training speakers overlap with those in development or evaluation sets for either corpus.}\par
To simulate a semi-supervised scenario, 10\% of the training speakers are randomly selected as the default labeled dataset, with the remainder treated as unlabeled data \textbf{ensuring no speaker overlap between the two datasets.} Specifically, 68 speakers are selected for DementiaBank Pitt and 36 speakers for JCCOCC MoCA, with investigators and participants each comprising half of the speakers in both labeled datasets. While transcriptions are unavailable for unlabeled data, speaker identities are assumed to be known, reflecting real-world settings where medical records typically include patient identifiers.
\par
\noindent
\textbf{Experiments setup:} We adopt Whisper-medium\footnote{https://huggingface.co/openai/whisper-medium} for its strong generalization ability, and utilize LoRA~\cite{lora} for finetuning\footnote{We apply LoRA to the ``query'', ``key", ``value", and ``att.out" of the attention module, with the rank set to 8 and alpha set to 16.}. The confidence score estimation module is a 3-layer residual feedforward DNN with a hidden dimension of 64, incorporating batch normalization and dropout, followed by a sigmoid gate. The speaker prompt length is set to 4, consistent with the best-performing speaker adaptation system in prior work~\cite{deng25_interspeech}. We construct the following systems for comparison: \textbf{1)} random sampling, which randomly selects a fixed percentile of utterances from the pseudo-labeled dataset and combines them with the labeled data for training; \textbf{2)} random partition incremental learning, which randomly partitions the pseudo-labeled data into subsets and progressively incorporates them in an incremental fashion. For all incremental pseudo-labeling systems, we uniformly divide the pseudo-labeled data into \textbf{5} equal-sized subsets and set the number of incremental steps to \textbf{5} accordingly, such that each incremental step incorporates an additional 20\% of the pseudo-labeled data.
\begin{table}[htbp]
    \vspace{-0.15cm}
    \caption{Performance comparison of different pseudo-labeling strategies on DementiaBank Pitt. ``Conf. Score" represents confidence score. ``Incre. Iter." denotes the number of incremental iterations. ``SA Relab." denotes speaker adaptive relabeling. Results of Sys. 7, 13, 19 are the mean $\pm$ standard deviation across 3 random seeds.}
    \centering
    \setlength\tabcolsep{1pt}
    \vspace{-0.2cm}
    \renewcommand{\arraystretch}{1.2}
    \resizebox{\linewidth}{!}
    {
    \begin{tabular}{c|c|c|c|c|c|c|cc|cc|c}
    \hline\hline 
    \multirow{3}{*}{Sys.} & 
    \multicolumn{5}{c|}{\shortstack{Pseudo Labeling}} & 
    \multirow{3}{*}{\shortstack{TTA}} & 
    \multicolumn{5}{c}{DementiaBank Pitt WER(\%)} \\
    \cline{2-6} \cline{8-12}
    &\multirow{2}{*}{\shortstack{Percentile of \\Data Used}}&\multirow{2}{*}{\shortstack{Data\\Selection}}&\multirow{2}{*}{\shortstack{\shortstack{Incre.\\ Iter.}}}&\multirow{2}{*}{\shortstack{SAT}}&\multirow{2}{*}{\shortstack{SA\\ relab.}}&&\multicolumn{2}{c|}{Dev.} & \multicolumn{2}{c|}{Eval.}&\multirow{2}{*}{All}\\
    \cline{8-11}
    &&&&&&&Par. &Inv. &Par. &Inv.\\
    \hline\hline
    {\cellcolor{yellow!25} 1} & \multicolumn{6}{c|}{Direct Finetune On 10\% Sup. Data} & 33.84 & 16.38 & 23.70 & 13.43 & 24.43 \\
    \hline
    {\cellcolor{green!10} 2} & \multirow{5}{*}{100\%} & \multirow{5}{*}{-} & - & \multicolumn{3}{c|}{\multirow{2}{*}{\xmark}}&32.03 & 15.57 & 23.62 & 13.32 & 23.36 \\
    
    3 & & &5 &\multicolumn{3}{c|}{}&30.98&14.74&22.63&15.43&22.51 \\
    \cline{4-7}
    4 & & &- &\multirow{3}{*}{\cmark}&\multirow{2}{*}{\xmark}&\multirow{3}{*}{\cmark}&32.06&15.34&22.84&13.98&23.17 \\
    \cline{4-4}
    5 & & &\multirow{2}{*}{5}&&&&30.46&14.48&22.34&14.32&22.12 \\
    \cline{6-6}
    6 & & &&&\cmark&&30.01&14.74&21.68&14.87&21.96 \\
    \hline
    7&\multirow{6}{*}{80\%}&Random&-&\multicolumn{3}{c|}{\multirow{3}{*}{\xmark}}&31.98$\pm$0.19&15.67$\pm$0.10&	22.98$\pm$0.16&	14.17$\pm$0.06&	23.30$\pm$0.19\\
    \cline{3-4}
    8&&\multirow{5}{*}{\shortstack{Conf.\\Score}}&-&\multicolumn{3}{c|}{}&31.64&15.07&22.40&13.54&22.81\\ 
    {\cellcolor{cyan!15}9}&&&4&\multicolumn{3}{c|}{}&30.85&14.91&22.23&14.87&22.45 \\ 
    \cline{4-12}
    10 & & &- &\multirow{3}{*}{\cmark}&\multirow{2}{*}{\xmark}&\multirow{3}{*}{\cmark}&31.50&14.99&22.13&12.87&22.66  \\
    \cline{4-4}
    {\cellcolor{orange!20} 11} & & &\multirow{2}{*}{4}&&&&30.26&14.58&21.50&13.32&21.91 \\
    \cline{6-6}
    12 & & &&&\cmark&&30.30&14.53&21.46&12.76&21.89 \\
    \hline
    
    13&\multirow{6}{*}{60\%}&Random&-&\multicolumn{3}{c|}{\multirow{3}{*}{\xmark}}&32.00$\pm$0.08&15.70$\pm$0.13&22.65$\pm$0.08&14.23$\pm$0.42&23.27$\pm$0.06\\
    \cline{3-4}
    14&&\multirow{5}{*}{\shortstack{Conf.\\Score}}&-&\multicolumn{3}{c|}{}&31.50&15.44&22.99&13.10&22.99 \\ 
    15&&&3&\multicolumn{3}{c|}{}&30.45&15.21&22.33&14.54&22.41 \\ 
    \cline{4-12}
    16 & & &- &\multirow{3}{*}{\cmark}&\multirow{2}{*}{\xmark}&\multirow{3}{*}{\cmark}&30.51&14.59&22.46&13.10&22.16   \\
    \cline{4-4}
    17 & & &\multirow{2}{*}{3}&&&&29.86&15.01&21.79&13.87&21.99 \\
    \cline{6-6}
    18 & & &&&\cmark&&30.17&14.68&21.54&13.54&21.93\\
    \hline
    
    19&\multirow{6}{*}{40\%}&Random&-&\multicolumn{3}{c|}{\multirow{3}{*}{\xmark}}&32.14$\pm$0.21&15.55$\pm$0.24&23.41$\pm$0.09&	13.76$\pm$0.38&23.38$\pm$0.04\\
    \cline{3-4}
    20&&\multirow{5}{*}{\shortstack{Conf.\\Score}}&-&\multicolumn{3}{c|}{}&32.21&15.20&23.49&13.54&23.27 \\ 
    21&&&2&\multicolumn{3}{c|}{}&31.36&15.66&22.82&14.21&23.03 \\ 
    \cline{4-12}
    22 & & &- &\multirow{3}{*}{\cmark}&\multirow{2}{*}{\xmark}&\multirow{3}{*}{\cmark}&31.44&14.97&22.91&13.54&22.78 \\
    \cline{4-4}
    23 & & &\multirow{2}{*}{2}&&&&30.68&15.47&22.23&13.65&22.56 \\
    \cline{6-6}
    24 & & &&&\cmark&&30.83&15.53&21.77&13.76& 22.58\\
    \hline    
\end{tabular}
}
\vspace{-0.2cm}
\label{tab:table_ablation}
\end{table}

\vspace{-0.5cm}
\begin{figure}[h]
    \centering        
    \includegraphics[width=0.28\textwidth]{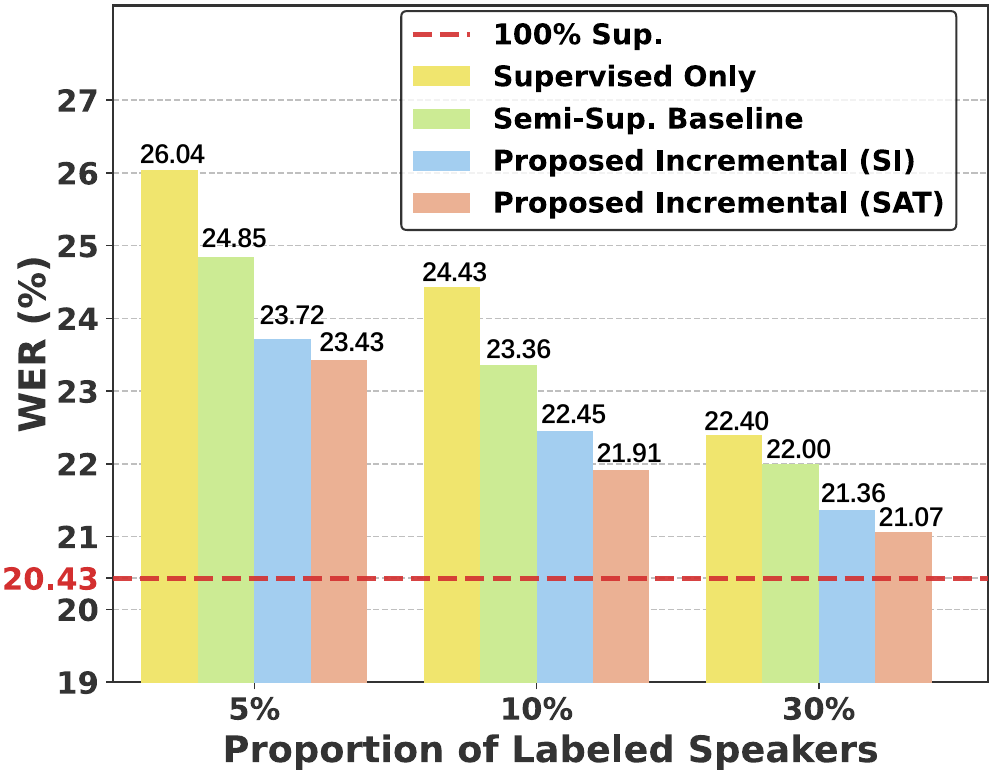}
    \vspace{-0.2cm}
    \caption{Performance of baseline systems and proposed methods across varying labeled speaker proportions.
}
\vspace{-0.1cm}
\label{fig:data_amount}
\end{figure}
\vspace{-0.3cm}

\begin{table}[htbp]
    \vspace{-0.3cm}
    \caption{Pseudo-label quality (WER/CER\%) comparison of different strategies on CEM ranked pseudo-labeled subsets. ``Incre.'' denotes incremental. ``10\% Sup.'' is the system trained on 10\% labeled data.}
    \centering
    \setlength\tabcolsep{1pt}
    \vspace{-0.2cm}
    \renewcommand{\arraystretch}{1.2}
    \resizebox{\linewidth}{!}
    {
    \begin{tabular}{c|c|c|c|c|c|c|c|c}
    \hline\hline 
    \multirow{2}{*}{Sys.} & 
    \multirow{2}{*}{Unlab. Data} & 
    \multicolumn{4}{c|}{DementiaBank Pitt WER (\%)} & 
    \multicolumn{3}{c}{JCCOCC MoCA CER (\%)} \\
    \cline{3-9}
    &&10\% Sup.&Incre. SI&Incre. SAT&Incre. SAT+SA relab.&10\% Sup.&Incre. SI&Incre. SAT\\
    \hline
    1&Rank1&6.67&6.67&6.67&6.67&2.42&2.42&2.42\\
    2&Rank2&12.00&11.97&11.98&11.93&4.67&4.55&4.52\\
    3&Rank3&19.98&19.47&19.54&19.36&9.03&8.57&7.69\\
    4&Rank4&34.21&31.94&31.34&31.01&13.71&12.71&12.53\\
    5&Rank5&63.77&56.12&55.38&55.10&25.56&23.18&19.59\\
    \hline\hline
\end{tabular}
}
\vspace{-0.3cm}
\label{pseudo_partition}
\end{table}

\par
\noindent
\textbf{Performance analysis:}
From the experimental results, several trends can be found:\\
\textbf{1)} Incorporating confidence score for data filtering achieves better performance compared with the semi-supervised baseline that trains on all pseudo-labeled data without filtering (Table \ref{tab:main_results}, Sys. 4 vs. 3). Furthermore, confidence-based filtering consistently outperforms random sampling across different data retention percentiles (Table \ref{tab:table_ablation}, Sys.8 vs. 7, Sys.14 vs. 13, Sys.20 vs. 19), confirming that utilizing confidence scores can effectively select high-quality pseudo-labels data. \\
\textbf{2)} Building upon confidence guided data ranking, the proposed incremental pseudo-labeling strategy consistently outperforms confidence-based filtering (Table \ref{tab:main_results}, Sys. 6 vs. 4). Unlike direct filtering, the proposed incremental strategy refines pseudo-label quality rather than excluding them entirely. Moreover, replacing confidence guided data ranking with random partitioning in the incremental framework leads to a clear performance degradation (Table \ref{tab:main_results}, Sys.5 vs. Sys.6), confirming that confidence-guided data ranking is essential to mitigating error accumulation across iterative training and validating the effectiveness of the proposed curriculum learning trajectory.\\
\textbf{3)} Integrating speaker modeling into the incremental pseudo-labeling framework yields further performance improvements (Table \ref{tab:main_results}, Sys.9 vs. Sys.6). This is consistent with the WER reduction when incorporating speaker modeling during incremental pseudo-labeling produces higher-quality pseudo-labels (Table \ref{pseudo_partition}, Incre. SAT+SA relab., Incre. SAT vs. Incre. SI) for both datasets. For systems employing only confidence-based filtering, introducing speaker modeling achieves consistent performance gains over those without speaker modeling across different data retention percentiles (Table \ref{tab:table_ablation}, Sys.10 vs. 8, Sys.16 vs. 14, Sys.22 vs. 20). \\
\textbf{4)} The proposed method achieves a steady WER reduction on the pseudo-labeled data, especially on the low confidence, poorer quality partition data (Table \ref{pseudo_partition}, Rank 3, 4, 5 data). This yields higher-quality pseudo-labels during iterative decoding. Thus, the proposed method outperforms the semi-supervised baseline by statistically significant WER or CER reductions of \textbf{1.45\%} and \textbf{2.27\%} absolute (\textbf{6.21\%} and \textbf{6.98\%} relative) (Table \ref{tab:main_results}, Sys.9 vs. 3). Furthermore, despite utilizing only 10\% of the labeled data, the proposed method achieves comparable results on the elderly participant evaluation set to the fully supervised systems (Table \ref{tab:main_results}, Sys.9 vs. 1).

\par
\noindent
\textbf{Ablation Studies:}
As shown in Table \ref{tab:table_ablation}, optimal results are obtained when retaining \textbf{80\%} of the pseudo-labeled data and performing \textbf{4} incremental iterations accordingly (Sys.11). The above settings are used on the JCCOCC MoCA dataset for the corresponding experiments in Table \ref{tab:main_results}. We further investigate the impact of varying labeled speaker proportions as illustrated in Fig. \ref{fig:data_amount}. The proposed incremental speaker adaptive pseudo-labeling method consistently achieves the best performance across different evaluated proportions of labeled speakers. Compared with the fully supervised upper bound, the WER degradation is only 0.64\% when using 30\% labeled speakers, and the proposed method can still achieve a WER of 23.43\% even under the extreme low-resource condition of only 5\% labeled speakers.

\vspace{-0.3cm}
\section{Conclusion}
This paper proposes a novel confidence score guided incremental and speaker adaptive pseudo-labeling approach for semi-supervised elderly speech recognition. An incremental strategy ensures the iterative refinement of confidence score guided pseudo-labels, while speaker modeling mitigates elderly speaker heterogeneity to generate higher-quality pseudo-labels. Experiments on the English DementiaBank Pitt and Cantonese JCCOCC MoCA elderly speech datasets suggest that the proposed method outperforms the semi-supervised baseline by statistically significant WER and CER reductions of 1.45\% and 2.27\% absolute (6.21\% and 6.98\% relative).

\section{Acknowledgements}
This research is supported by Hong Kong RGC GRF grant No. 
14200021, 14200324, 
Basic Research Project of Institute of Software, Chinese Academy of Sciences ISCAS-JCMS-202306, and Youth Innovation Promotion Association CAS Grant 2023119.

\section{Generative AI Use Disclosure}
Generative AI tools were used only for language editing and improving readability during the preparation of this manuscript. These tools were not used to generate core scientific ideas, experimental data, or technical contributions. All authors have thoroughly reviewed and approved the final manuscript and assume full responsibility for the integrity of its entire content.

\bibliographystyle{IEEEtran}
\bibliography{mybib}

\end{document}